\documentclass[twocolumn,english,prb]{revtex4-1}
\usepackage{hyperref}
\usepackage{amsmath}
\usepackage{graphicx}
\usepackage[caption=false]{subfig}

\renewcommand{\Im}{\operatorname{Im}}

\begin{document}

\title{Spin correlation functions and decay of quasiparticles
 in  XXZ spin chain at $T >0$.}
\author{Igor Poboiko}
\affiliation{L. D. Landau Institute for Theoretical Physics, 117940 Moscow, Russia}
\affiliation{Moscow Institute for Physics and Technology, Dolgoprudny, Moscow region, Russia}
\author{Mikhail Feigel'man}
\affiliation{L. D. Landau Institute for Theoretical Physics, 117940 Moscow, Russia}
\affiliation{Moscow Institute for Physics and Technology, Dolgoprudny, Moscow region, Russia}
\date{\today}

\begin{abstract}
We study one-dimensional anisotropic XXZ spin-$\frac12$  model with ferromagnetic sign of the coupling
and $z-z$ exchange constant  $J_z = \Delta J$, where $\Delta < 1$, and  $J$ is the coupling within XY
spin plane. We calculate damping of low-energy excitations with $\omega \ll T $ due to their
scattering from thermal excitation bath with temperature $T \ll J$, taking into account nonzero 
curvature  of the excitation spectrum, $\epsilon(q) = u q + \delta\epsilon(q)$. 
We calculate also  longitudinal spin-spin correlation function 
$\langle S^z(x,t)S^z(0,0) \rangle$  at $x \approx ut$  and find the shape of the spreading 
"wave packet".
\end{abstract}

\maketitle
\section{Introduction}

Particle and heat transport in disordered spin chains is actively studied  last years~\cite{MBL1,MBL2,MBL3,MBL4,MBL5,MBL6,MBL7,MBL8} 
since these models present apparently simple examples of interacting quantum systems with frozen disorder.
One particular example of a such  studies is presented by  our recent work~\cite{PF2015} where thermal
 conductivity of weakly disordered spin-$\frac12$ anisotropic (XXZ) chain was calculated at low temperatures
 $T \ll J$ and in the anisotropy range  $ 1/2 < \Delta = J^{zz}/J  < 1$.  In that study we employed Jordan-Wigner
transformation to the Fermionic chain problem and then used standard Bozonization technique~\cite{GiamarchiBook}  
which is limited to the  Luttinger Liquid approximation with a  linear spectrum (LLL).  
It was found in~\cite{PF2015} that such an approximation is valid for sufficiently low temperatures if
  $\Delta < \cos\frac{\pi}{5} \approx 0.81$ and frozen disorder is weak, $\langle h^2\rangle \ll J^2$;
 on the other hand, LLL approximation breaks down completely at larger values of $\Delta$.  

Effects of nonlinear spectrum in 1D quantum liquids have been actively studied from different viewpoints, 
see recent review~\cite{Adilet} and papers~\cite{Samokhin,Aristov,Gangardt,Protopopov}.
 There are basically two possible approaches to the problem, using either
Fermionic or Bosonic representation. In the Fermionic representation, bare  spectrum of 
 quasiparticles is $\epsilon(q) = u q + \delta\epsilon(q)$,  with $\delta\epsilon(q) \ll uq$ at small $q$.
For a vanishing or very weak interaction between Fermions, one can easily construct Bosonic correlation function 
as composite objects made out of pair of Fermions. Then spectral function $g(\omega,q)$ corresponding to
 any Bosonic operator  can be obtained as a convolution of two Fermionic spectral functions
and thus will have a finite width $\sim \delta\epsilon(q)$.   However, this approach becomes rather complicated
if interaction between Fermions is not weak.

On the other hand, strong Fermion-Fermion interaction can be conviniently described in the Bosonic
representation, but then nonlinearity of the original Fermionic spectrum transforms into non-linear
interaction between LL bosons. This interaction is cubic if $\delta\epsilon(q) \propto q^2$  (which is a general case)
 and quartic in the case of particle-hole symmetry when $\delta\epsilon(q) \propto q^3$.
 The problem  with treatment of such an interaction  within LL theory is nontrivial
 since for 1-dimensional particles with linear spectrum  all interactions between particles with the same
 sign  of velocity are of resonant nature as absolute value of velocity is fixed and particles stay in contact
for arbitrarily long time.  As a result, straightforward second-order perturbation theory  produces diverging 
result for the imaginary part of the self-energy at the "light cone":   
$\Im\Sigma(\omega, q) \propto \delta(\omega - uq)$.

In the present paper we develop diagrammatic approach to the calculation of $\Im\Sigma(\omega \approx u q)$ for
 Bosonic excitations with energy $\omega$ which is low in comparison with the temperature $T$. 
 We show that in the limit $\omega \ll T$ quasiparticle decay rate $\Gamma(\omega, q)$
 can be found self-consistently near the light-cone, leading to a finite (although non-analytic) result.
 We consider below two different situations: pure exchange spin chain Hamiltonian symmetric w.r.t. to 
$S^z \to -S^z$ inversion, and spin chain in presence of uniform magnetic field $h$ along $z$ axis, 
with $h \gg T $. In the latter case our result is very similar to the one obtained long ago 
by A.F.Andreev~\cite{Andreev}  for fluctuational correction to viscosity in a 1-dimensional
classical hydrodynamics and rederived in Ref.~\cite{Samokhin}.  We are not aware of any previous
 calculations,   which accounts for spectrum nonlinearity, for the symmetric case $h=0$.

We emphasize that the decay rate of bosonic excitations $\Gamma(\omega)$  calculated in this paper does not coinside
with inelastic  relaxation rate;  rather it provides the measure  of coherence for the specific type
of excitations  we study. The meaning of $\Gamma(\omega)$ is that it determines (due to direct local relation
between bosonic and spin variables)  the shape of the dynamic
spin-spin correlation function $\langle S^z(x,t)S^z(0,0) \rangle$ at $x \approx \pm ut$.

For the possibility to develop a self-consistent diagrammatic approach, the  condition
$\omega \ll T$ is crucial.  In the opposite limit more involved calculations are needed,  like those
developed by Imambekov \textit{et al}~\cite{Adilet} for the shape of bosonic spectral function
at $T=0$. 

Another piece of work we  mention is related with a discussion of the presence of
the Drude weight in these spin chains~\cite{Affleck2011}.   This interesting issue is beyond the
scope of our study since it is related to the decay  rate of the uniform current $\Gamma(\omega,0)$,
whereas  we calculated $\Gamma(\omega,q)$ for $\omega \approx uq$.

The rest of the paper is organized as follows: we formulate our model in Sec. \ref{sec:Model}, develop 
diagrammatic perturbation theory in Sec.\ref{sec:LowestOrder}  and extend it to the self-consistent  approach in Sec.\ref{sec:SelfConsistent},
where major results for the decay rates $\Gamma$ are obtained. Sec.\ref{sec:CorrelationFunctions} is devoted to the calculation of dynamic
spin-spin correlation function $\langle S^z(x,t)S^z(0,0) \rangle$  at $x \approx \pm ut$ which describe
spreading of the excitation wave-packet due to scattering on thermal excitations; finally, Sec.\ref{sec:Conclusions} contains our
conclusions.

\section{The model}
\label{sec:Model}

We study spin-$\frac{1}{2}$ anisotropic XXZ spin chain with external magnetic field applied along the $z$ direction, which is described by the following Hamiltonian:
\begin{equation}
	\label{eq:Hspin}
	\hat{H} = -J \sum_n \left(\hat{S}_n^x \hat{S}_{n+1}^x + \hat{S}_n^y \hat{S}_{n+1}^y + 
	\Delta \hat{S}_n^z \hat{S}_{n+1}^z + \frac{h}{J} \hat{S}_n^z \right)
\end{equation}
The sign of exchange constant in the $XY$ plane can be changed utilizing the canonical transformation $\hat{S}_{n}^{x}\mapsto(-1)^{n}\hat{S}_{n}^{x}$, $\hat{S}_{n}^{y}\mapsto(-1)^{n}\hat{S}_{n}^{y}$, and $\hat{S}_{n}^{z}\mapsto\hat{S}_{n}^{z}$; this allows us to fix the sign of coupling constant $J > 0$. Positive value of $\Delta$ corresponds to ferromagnetic exchange, while negative sign corresponds to antiferromagnetic exchange.

By means of the Jordan-Wigner transformation, the Hamiltonian \eqref{eq:Hspin} is equivalent to the following Hamiltonian of interacting spinless fermions:
\begin{equation}
	\label{eq:HJW}
	\hat{H}=-J\sum_{n}\left(\frac{1}{2}c_{n}^{\dagger}c_{n+1}+h.c. + \Delta \rho_n \rho_{n+1} + \frac{h}{J} \rho_n\right),
\end{equation}
with fermion density operator $\rho_{n}=c_{n}^{\dagger}c_{n}-\frac{1}{2}\equiv\hat{S}_{n}^{z}$.

For small compared to the bandwidth magnetic field $h \ll J$ and values of anisotropy parameter $-1 < \Delta < 1$, the low-energy properties the \eqref{eq:HJW} is described by the Luttinger Liquid model \cite{GiamarchiBook}. 
The model describes fermion density excitations, which is related to the bosonic field $\phi(x)$ as $\rho(x) = -\frac{1}{\pi} \partial_x \phi(x)$. After introducing canonical conjugate momentum $\Pi(x)$, so that $[\phi(x), \Pi(y)] = i \delta(x-y)$, the quadratic part of the Hamiltonian density, including its field-dependent part, are written as follows:
\begin{equation}
\label{eq:HLL}
\hat{\mathcal{H}}_{0}=\frac{1}{2\pi}\left(\frac{u}{K}(\partial_{x}\phi)^{2}+uK(\pi\Pi)^{2}\right),\quad \hat{\mathcal{H}}_h = - \frac{1}{\pi} h \partial_x \phi 
\end{equation}
with $u$ being plasmon group velocity and $K$ being dimemsionless Luttinger parameter, whose values are expressed in terms of coupling constant $J$, lattice constant $a$ and anisotropy parameter $\Delta$ as follows:
\begin{equation}
\Delta=\cos\frac{\pi}{2K}, \qquad u=\frac{Ja}{2}\frac{\sin\frac{\pi}{2K}}{1-\frac{1}{2K}}
\end{equation}

If one considers the quadratic Hamiltonian only, one immediately obtains that magnetic field term can be easily discarded using simple phase shift $\partial_x \phi \mapsto \partial_x \phi + \frac{K}{u} h$. However, it is not the case if one takes into the account irrelevant terms, that are higher order in bosonic fields.

Alternatively, one can introduce densities of right- and left-moving fermions $R(x)$ and $L(x)$, consisting of Fourier harmonics of $\rho(x)$ with $k > 0$ and $k < 0$ respectively. For further calculations, it will be convenient to rescale them with $\sqrt{K}$ factor, so that $\rho(x) = \sqrt{K} (R(x) + L(x))$. Explicit expression for the densities is $R(x),L(x) = -\frac{1}{2 \pi \sqrt{K}} \partial_x \phi(x) \pm \frac{\sqrt{K}}{2} \Pi(x)$, with upper sign corresponding to the right-movers and lower sign corresponding to the left-movers. In terms of these fields, the quadratic part of the Hamiltonian density is written as follows:
\begin{equation}
\hat{\mathcal{H}}_{0}=\pi u(R^{2}+L^{2}),\quad \hat{\mathcal{H}}_{h} = h\sqrt{K}(R+L)
\end{equation}

In addition to the quadratic part, there are also irrelevant in the RG sense terms in the Hamiltonian, which keep information about the lattice nature of the original model \cite{Lukyanov}. The two most important of those irrelevant operators are so-called ``umklapp term'' and the ``band curvature term''.

The ``umklapp term'' is written as follows:
\begin{equation}
\hat{\mathcal{H}}_u = \lambda\frac{u}{a^2} \cos (4\phi(x)),
\end{equation}
Such term has scaling dimension $2-4K$ and is irrelevant at $K > \frac{1}{2}$, that is $\Delta > -1$. 
The effect of such term at nonzero temperatures was studied in Ref. \cite{Affleck2011}, where it was shown that it leads to the finite decay of the quasiparticles $\Gamma \propto \lambda^2 T^{8 K - 3}$, which is small at sufficiently large $K$.

It is convenient to write the ``band curvature term'' in terms of $R(x)$ and $L(x)$. Corresponding expression reads:
\begin{equation}
\label{eq:Hbc4}
\hat{\mathcal H}^{(4)}_{b.c.}=-\frac{\alpha}{2}(\lambda_{+}R^{2}L^{2}+\lambda_{-}(R^{4}+L^{4})),
\end{equation}
with parameters
\begin{equation}
\label{eq:alpha}
\alpha = 4\pi^{3}ua^{2}
\end{equation}
\begin{equation}
\lambda_+ = \frac{1}{2\pi}\tan\frac{\pi K}{2K-1},
\end{equation}
\begin{equation}
\lambda_- = \frac{1}{24\pi K}\frac{\Gamma\left(\frac{3K}{2K-1}\right)}{\Gamma\left(\frac{3}{4K-2}\right)}\frac{\Gamma^{3}\left(\frac{1}{4K-2}\right)}{\Gamma^{3}\left(\frac{K}{2K-1}\right)}.
\end{equation}
These terms are also irrelevant with negative scaling dimension $-2$; however, at finite temperatures they become more important compared to the umklapp term at $K > 1$, that is $\Delta > 0$. 
Thus at zero magnetic field we will restrict ourselves to the region $K > 1$ and consider Hamiltonian $\hat{\mathcal{H}} = \hat{\mathcal{H}}_0 + \hat{\mathcal{H}}_{b.c.}^{(4)}$.

Magnetic field behaves as a chemical potential for the JW fermions. Shifting it away from the half-filling point with particle-hole symmetry leads to the appearing of the quadratic terms in the quasiparticle dispersion $\delta\epsilon(q) \propto q^2$, which in language of the bosonization corresponds to the cubic interaction terms.
These terms naturally appear if one performs the phase shift $R,L \mapsto R,L - \frac{\sqrt{K}}{2 \pi u} h$, that discards the magnetic field term from the quadratic part of the Hamiltonian \eqref{eq:HLL} and takes into the account the nonlinear terms \eqref{eq:Hbc4}; they read as follows:
\begin{equation}
\label{eq:Hbc3}
{\cal H}_{b.c.}^{(3)}=\frac{\alpha_{1}}{3}(R^{3}+L^{3})+\frac{\alpha_{2}}{2}(R^{2}L+RL^{2}),
\end{equation}
with constants
\begin{equation}
\alpha_{1}=\frac{3\alpha\sqrt{K}\lambda_{-}h}{\pi u},\quad \alpha_{2}=\frac{\alpha\sqrt{K}\lambda_{+}h}{\pi u}.
\end{equation}
The effect of quadratic curvature on the transport properties was studied in Refs. \cite{Aristov,Samokhin}. At sufficiently small magnetic fields, cubic terms are less important compared to the quartic terms; below it will be shown that crossover to the regime where cubic terms dominate in the quasiparticle decay happens at $h \sim T$. The total Hamiltonian at nonzero magnetic fields is thus  $\hat{\mathcal{H}} = \hat{\mathcal{H}}_0 + \hat{\mathcal{H}}_{b.c.}^{(3)} + \hat{\mathcal{H}}_{b.c.}^{(4)}$, while the latter term can be neglected when the magnetic field is large enough.

Below we will study only the effect of ``band curvature term'' on the decay of the quasiparticles, restricting ourselves to the region
$K > 1$  where it is  definitely more important at low $T$, compared to the ``umklapp term''.

\section{Perturbation theory}
In this section we will develop a perturbation theory with respect to nonlinear terms in the Hamiltonian. To begin with, we introduce retarded correlation functions (here we use short notation $\textbf{x} = (x,t)$ and $\textbf{q} = (q,\omega)$:
\begin{eqnarray}
g_{ret}^{(R)}(\mathbf{x}_{1}-\mathbf{x}_{2})&=-i\theta(t_1-t_2)\left\langle \left[R(\mathbf{x}_1),R(\mathbf{x}_2)\right]\right\rangle \\
g_{ret}^{(L)}(\mathbf{x}_{1}-\mathbf{x}_{2})&=-i\theta(t_1-t_2)\left\langle \left[L(\mathbf{x}_{1}),L(\mathbf{x}_{2})\right]\right\rangle
\end{eqnarray}
with following unperturbed values calculated with respect to $\hat{\mathcal{H}}_0$:
\begin{equation}
\label{eq:ZeroGreenFunction}
g_{ret}^{(R/L,0)}(\mathbf{q})=\pm\frac{q}{2\pi}\frac{1}{\omega + i0\mp uq}
\end{equation}
in which upper sign corresponds to $R$ and lower sign corresponds to $L$. 

To sum up reducible Feynman diagrams in perturbation theory series we will use standard Dyson equation. The equation and its solution are written as follows (for both right and left movers):
\begin{equation}
g_{ret}=g_{ret}^{(0)}+g_{ret}^{(0)}\Sigma_{ret}g_{ret}
\end{equation}
\begin{equation}
g_{ret}^{(R/L)}(\mathbf{q})=\pm\frac{q}{2\pi}\frac{1}{\omega - u q+i0\mp\frac{q}{2\pi}\Sigma_{ret}^{(R/L)}(\mathbf{q})}
\end{equation}

We are interested in the decay rate of the quasiparticles, that is governed by the imaginary part of the self-energy. Neglecting the renormalization of the quasiparticle spectra and wavefunction amplitude, that correspond to its real part, one immediately obtains the following form of the Greens function:
\begin{equation}
\label{eq:DressedGreenFunction}
g_{ret}^{(R/L)}(\mathbf{q})=\pm\frac{q}{2\pi}\frac{1}{\omega \mp u q +i\Gamma^{(R/L)}(\mathbf{q})}
\end{equation}
\begin{equation}
\label{eq:DecayRate}
\Gamma^{(R/L)}(\mathbf{q}) = \mp\frac{q}{2\pi}\Im\Sigma_{ret}^{(R/L)}(\mathbf{q})
\end{equation}

In the calculations below we will use Keldysh diagram technique for nonzero temperatures. This technique introduces additional Keldysh space. Green functions are 2 by 2 matrices in Keldysh space with following structure:
\begin{equation}
\hat{g}=\begin{pmatrix}g_{K} & g_{ret}\\
g_{adv} & 0
\end{pmatrix}
\end{equation}
The Keldysh component of the Green functions is expressed via the probability function $f(\omega) = \coth \frac{\omega}{2 T}$ as follows:
\begin{equation}
\label{eq:KeldyshEquilibrium}
g_K(\omega) = f(\omega) (g_{ret}(\omega) - g_{adv}(\omega)) = 2 i f(\omega) {\rm Im} g_{ret} (\omega)
\end{equation}
Interaction vertices in Keldysh technique has symmetric tensor structure with following components (we use the standard notation of ``classical'' and ``quantum'' fields as it is introduced e.g. in Ref. \cite{KamenevLevchenko}, so indices $a,b,c \in \{cl, q\}$):
\begin{equation}
\hat{\gamma}_{q,q,q}^{(3R)}=\hat{\gamma}_{cl,q,q}^{(3R)}=-\sqrt{2}\alpha_{1},\quad\hat{\gamma}_{abc}^{(3L)}\equiv\hat{\gamma}_{abc}^{(3R)},
\end{equation}
\begin{equation}
\hat{\gamma}_{q,q,q}^{(2R,L)}=\hat{\gamma}_{cl,q,q}^{(2R,L)}=-\frac{1}{\sqrt{2}}\alpha_{2},\quad\hat{\gamma}_{abc}^{(R,2L)}\equiv\hat{\gamma}_{abc}^{(2R,L)},
\end{equation}
\begin{equation}
\hat{\gamma}_{cl,cl,cl,q}^{(4R)}=\hat{\gamma}_{cl,q,q,q}^{(4R)}=6\alpha\lambda_{-},\quad\hat{\gamma}_{abcd}^{(4L)}\equiv\hat{\gamma}_{abcd}^{(4R)},
\end{equation}
\begin{equation}
\hat{\gamma}_{cl,cl,cl,q}^{(2R,2L)}=\hat{\gamma}_{cl,q,q,q}^{(2R,2L)}=\alpha\lambda_{+},
\end{equation}

Below we will focus only on the calculation of retarded self-energy part for density of right-moving particles; expressions for left-movers can be extracted straightforwardly by replacing $q \mapsto -q$ in $\Sigma^{(R)}_{ret}(\textbf{q})$ due to L-R symmetry.

\subsection{Lowest-order calculation}
\label{sec:LowestOrder}

\subsubsection{Zero magnetic field}
\begin{figure}
	\subfloat[$\Sigma^{(R \to 3R)}_{ret}$]{\includegraphics[width=0.45\columnwidth]{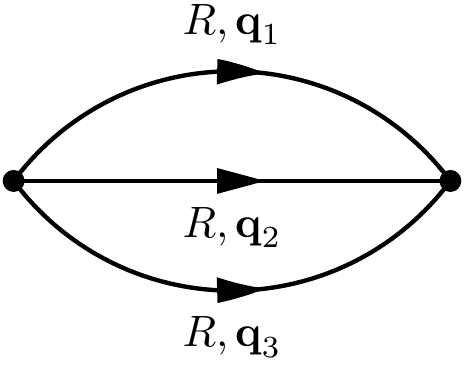}}	
	\subfloat[$\Sigma^{(R \to R+2L)}_{ret}$]{\includegraphics[width=0.45\columnwidth]{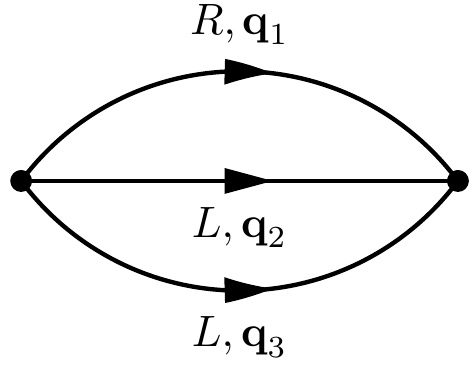}}	
	\caption{Diagrams for self-energy $\Sigma_{ret}^{(R)}$ coming from the Hamiltonian \eqref{eq:Hbc4}. Lines correspond to Green functions, that are $2 \times 2$ matrices in Keldysh space; while vertices have tensor structure in Keldysh space}
	\label{fig:selfenergy4}
\end{figure}
In the second order of perturbation theory, there are two diagrams for retarded self-energy of right-movers coming from terms $R^4$ and $R^2 L^2$ in \eqref{eq:Hbc4}, see Fig. \ref{fig:selfenergy4}. We will focus on the imaginary part of the self-energy, since it governs the decay of the quasiparticles.

It is possible to express the contribution to the imaginary part of self-energy coming from the first diagram via the spectral weights $\Im g_{ret}^{(R,L)}(\mathbf{q})$:
\begin{multline}
\label{eq:ImSigmaRto3R}
{\rm Im}\Sigma_{ret}^{(R\to 3R)}(\mathbf{q})=\frac{3\alpha^{2}\lambda_{-}^{2}}{2\pi^{4}}\int d^{2}\mathbf{q}_{1}d^{2}\mathbf{q}_{2}\times\\
\times{\rm Im}g_{ret}^{(R)}(\mathbf{q}_{1}){\rm Im}g_{ret}^{(R)}(\mathbf{q}_{2}){\rm Im}g_{ret}^{(R)}(\mathbf{q}_{3})\times\\
\times(1+f(\omega_{2})f(\omega_{3})+f(\omega_{1})f(\omega_{3})+f(\omega_{1})f(\omega_{2}))
\end{multline}

The unperturbed spectral weights are delta-peaked on the mass shell $\Im g_{ret}^{(R,L)}(\mathbf{q}) \propto \delta(\omega \mp u q)$. This fact together with energy and momentum conservation laws lead to singular behavior of self-energy, which is also shows delta-peak behavior on the mass shell:
\begin{multline}
\label{eq:ImSigmaRto3Rresult}
{\rm Im}\Sigma_{ret}^{(R\to3R)}(\mathbf{q})=\\
=-\frac{\alpha^{2}\lambda_{-}^{2}}{160\pi^{4}}\delta(\omega-uq)q\left[q^{2}+\left(\frac{2\pi T}{u}\right)^{2}\right]\left[q^{2}+4\left(\frac{2\pi T}{u}\right)^{2}\right].
\end{multline}
Singularity on the mass shell in \eqref{eq:ImSigmaRto3R} is an artifact of the lowest-order perturbation  theory calculation;  below we will see that its proper regularization leads to a behavior that is 
finite everywhere and non-analytic as function of the coupling constant  $\alpha \lambda_-$.

In order to obtain meaningful result for this diagram, one should perform the calculation of \eqref{eq:ImSigmaRto3R} self-consistently by putting ``dressed'' Greens functions instead of bare ones: it is known that the exact spectral weight calculated with ``dressed'' Green functions have nonzero width and finite height \cite{Adilet}. Such procedure is equivalent to resummation of an infinite series of Feynman diagrams that are most singular near the mass shell. Below, in Sec. \ref{sec:SelfConsistent}, we will perform such self-consistent procedure.

The analytic expression for the second diagram from Fig. \ref{fig:selfenergy4} calculated with bare Greens functions can only be expressed via the polylogarithm function; however, there are three cases of interest, namely zero temperature $T = 0$, large temperature $T \gg \omega, uq$ and on the mass shell $\omega = uq$, where the asymptotic behavior can be obtained. At zero temperature it yields:
\begin{multline}
\label{eq:ImSigmaRtoRplus2Lresult1}
{\rm Im}\Sigma_{ret}^{(R\to R+2L)}(\mathbf{q}) \approx\\
\approx -\frac{\alpha^2\lambda_+^2}{3072\pi^{4} u^{5}}(\omega-uq)(\omega+uq)^{3}\theta(\omega^{2}-u^{2}q^{2}){\rm sign}\omega,
\end{multline}
while both for cases $T \gg \omega,uq$ and $\omega = u q$ it yields:
\begin{equation}
\label{eq:ImSigmaRtoRplus2Lresult2}
{\rm Im}\Sigma_{ret}^{(R\to R+2L)}(\mathbf{q})\approx-\frac{\alpha^{2}\lambda_{+}^{2}}{48\pi^{2} u^{5}}T^{3}\omega
\end{equation}

According to the equation \eqref{eq:DecayRate}, this corresponds to quasiparticle decay rate $\Gamma(\omega = u q) \propto \omega^{2}T^{3}/J^{4}$. As we will see later, such decay rate is negligible compared to the singular contribution coming from first diagram.

\subsubsection{Nonzero magnetic field}
\begin{figure}
	\subfloat[$\Sigma^{(R \to 2R)}_{ret}$]{\includegraphics[width=0.45\columnwidth]{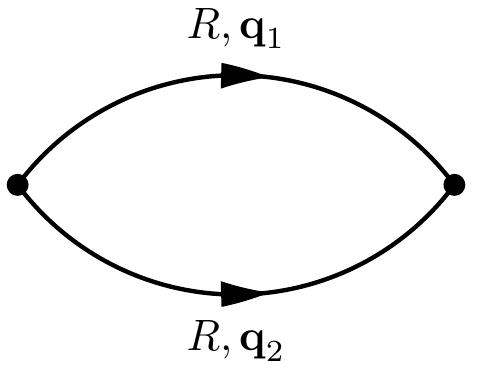}}	
	\subfloat[$\Sigma^{(R \to R+L)}_{ret}$]{\includegraphics[width=0.45\columnwidth]{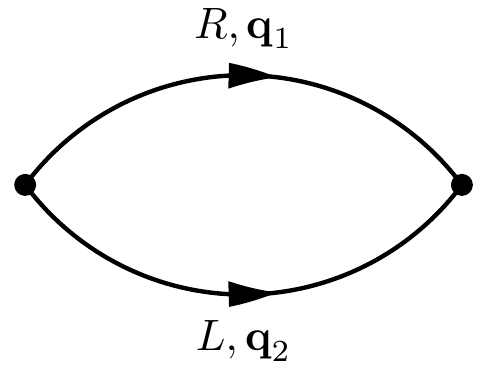}}	
	\caption{Diagrams for self-energy $\Sigma_{ret}^{(R)}$ coming from the Hamiltonian \eqref{eq:Hbc3} in the presence of nonzero magnetic field.}
	\label{fig:selfenergy3}
\end{figure}

Here we perform the same calculations as above, but with respect to the perturbation \eqref{eq:Hbc3}. There are two Feynman diagrams giving contribution to the imaginary part of self-energy (see Fig. \ref{fig:selfenergy3}). 

The expression for the first diagram in terms of the spectral weights $\Im g_{ret}^{(R)}(\mathbf{q})$ is following:
\begin{multline}
\label{eq:ImSigmaRto2R}
{\rm Im}\Sigma_{ret}^{(R\to2R)}(\mathbf{q})=-\frac{\alpha_{1}^{2}}{2\pi^{2}}\int d^{2}\mathbf{q}_{1}\times\\
\times{\rm Im}g_{ret}^{(R)}(\mathbf{q}_{1}){\rm Im}g_{ret}^{(R)}(\mathbf{q}_{2})(f(\omega_{1})+f(\omega_{2}))
\end{multline}

Direct calculation (see Appendix \ref{sec:AppendixImSigma}) immediately recovers the same delta-peaked behavior on the mass shell due to the very same reason as in previous case:
\begin{equation}
\label{eq:ImSigmaRto2Rresult}
{\rm Im}\Sigma_{ret}^{(R\to2R)}=-\frac{\alpha_{1}^{2}}{24\pi^{2}}\delta(\omega-uq)q\left(q^{2}+\left(\frac{2\pi T}{u}\right)^{2}\right)
\end{equation}

Similarly, the proper way to deal with it is to solve the equation \eqref{eq:ImSigmaRto2R} self-consistently utilizing the ``dressed'' Greens functions instead of bare ones.

The second diagram yields:
\begin{multline}
\label{eq:ImSigmaRtoRplusLresult}
{\rm Im}\Sigma_{ret}^{(R\to R+L)}(\mathbf{q})=-\frac{\alpha_{2}^{2}}{128\pi^{2}u^{3}}(\omega^{2}-u^{2}q^{2})\times\\
\times\left(f\left(\frac{\omega-uq}{2}\right)+f\left(\frac{\omega+uq}{2}\right)\right)
\end{multline}
The same asymptotic analysis of cases $T = 0$, $T \gg \omega,uq$ and $\omega = uq$ as for quartic interaction can be performed, yielding result for zero temperature:
\begin{equation}
{\rm Im}\Sigma_{ret}^{(R\to R+L)}(\mathbf{q})\approx-\frac{\alpha_{2}^{2}}{64\pi^{2}u^{3}}(\omega^{2}-u^{2}q^{2})\theta(\omega^{2}-u^{2}q^{2}){\rm sign}\omega,
\end{equation}
and both for $T \gg \omega,uq$ and $\omega = uq$:
\begin{equation}
{\rm Im}\Sigma_{ret}^{(R\to R+L)}(\mathbf{q})=-\frac{\alpha_{2}^{2}}{16\pi^{2}u^{3}}T\omega
\end{equation}

According to \eqref{eq:DecayRate}, this leads to the decay rate $\Gamma(\omega) \propto h^2 \omega^2 T / J^4$. However, it is to be stressed again that such decay rate is negligible compared to the singular one coming from the first diagram and Equation \eqref{eq:ImSigmaRto2R}.

Cubic vertices $R^2 L$ and $R L^2$ also lead to the appearance of the offdiagonal Green functions, such as $\left<R(\mathbf{x}_1) L(\mathbf{x}_2)\right>_{ret}$ \cite{Aristov}. They will manifest themselves e.g. when calculating spin correlation functions, see Sec. \ref{sec:CorrelationFunctions}; however, they contain second power of ``coupling constant'' $\alpha_2$ and thus are parametrically smaller compared to the contribution coming from diagonal terms.

\subsection{Self-consistent calculation}
\label{sec:SelfConsistent}
We now switch to the self-consistent calculation as it was outlined above in Sec. \ref{sec:LowestOrder}. The main contribution for linewidth $\Gamma$ near the mass shell $\omega = u q$ comes from \eqref{eq:ImSigmaRto3R} and \eqref{eq:ImSigmaRto2R}; for now we will neglect contributions coming from other processes. We will also neglect the renormalization of quasiparticle spectra due to temperature effects and assume the dressed Green functions taking the form \eqref{eq:DressedGreenFunction}.
The cases of zero and nonzero magnetic fields will be studied separately below.

\subsubsection{Zero magnetic field}
We substitute Green functions of the form \eqref{eq:DressedGreenFunction} to the Eq.\eqref{eq:ImSigmaRto3R}, and introduce new variable: instead of $q$ we will use deviation from mass shell defined as $\epsilon = \omega - u q$. Since we are interested in well-defined quasiparticles $\omega, u q \gg \Gamma(\omega,q)$, we will also replace $q/2\pi$ prefactor in \eqref{eq:DressedGreenFunction} by $\omega / 2 \pi u$. After doing so, we arrive at the following integral equation for $\Gamma(\omega, \epsilon)$:
\begin{multline}
\Gamma(\omega,\epsilon)=\frac{3\alpha^{2}\lambda_{-}^{2}}{32\pi^{5}u^{5}}q\int d\omega_i\omega_{1}\omega_{2}\omega_{3} J_1(\epsilon, \omega_i)\times\\
\times(1+f(\omega_{2})f(\omega_{3})+f(\omega_{1})f(\omega_{3})+f(\omega_{1})f(\omega_{2}))
\end{multline}
\begin{multline}
J_1(\epsilon, \omega_i) = \int d\epsilon_i\frac{1}{\pi^3} \frac{\Gamma(\omega_1,\epsilon_1)}{\epsilon_{1}^{2}+\Gamma^{2}(\omega_{1},\epsilon_{1})} \frac{\Gamma(\omega_{2},\epsilon_{2})}{\epsilon_{2}^{2}+\Gamma^{2}(\omega_{2},\epsilon_{2})}\times\\
\times\frac{\Gamma(\omega_{3},\epsilon_{3})}{\epsilon_{3}^{2}+\Gamma^{2}(\omega_{3},\epsilon_{3})}
\end{multline}
Integration is performed over all $\omega_i$ and $\epsilon_i$ taking into account energy-momentum conservation $\omega_1 + \omega_2 + \omega_3 = \omega$ and $\epsilon_1 + \epsilon_2 + \epsilon_3 = \epsilon$.

All the $\epsilon_i$ in the $J_1$ integral are typically of the order of $\Gamma(\omega) \equiv \Gamma(\omega, \epsilon=0)$.
To start our analysis,  we will assume that $\Gamma(\omega,\epsilon \sim \Gamma(\omega)) \approx \mathrm{const}$, which will allow us to neglect $\epsilon$-dependence and perform integration over $\epsilon_i$ explicitly (this assumption will be checked \textit{a posteriori}).
Similarly, we will put $\epsilon = 0$ in the right-hand side of the equation and obtain a closed equation for $\Gamma(\omega)$. 
 This calculation yields:
\begin{equation}
J_1(\epsilon, \omega_i) = \frac{1}{\pi}\frac{\Gamma(\omega_{1})+\Gamma(\omega_{2})+\Gamma(\omega_{3})}{\epsilon^{2}+(\Gamma(\omega_{1})+\Gamma(\omega_{2})+\Gamma(\omega_{3}))^{2}}
\end{equation}
Next we introduce dimensionless linewidth $\Gamma(\omega)=\frac{\alpha\lambda_{-}}{u^{3}}T\omega^{2}\cdot\gamma(z \equiv \frac{\omega}{T})$. The corresponding equation for $\gamma(z)$ reads as follows:
\begin{multline}
\gamma(z)=\frac{3}{32\pi^{6}}\cdot\frac{1}{z}\int dz_i\frac{z_{1}z_{2}z_{3}}{\gamma(z_{1})z_{1}^{2}+\gamma(z_{2})z_{2}^{2}+\gamma(z_{3})z_{3}^{2}}\times\\
\times(1+f(z_{2})f(z_{3})+f(z_{1})f(z_{3})+f(z_{1})f(z_{2}))
\end{multline}
If one assumes slow dependence of $\gamma(z)$ on $z$ for small $z$ in the right-hand side of the equation, one immediately obtains that integral is logarithmic in $z$, and the main contribution comes from $z \lesssim z_i \lesssim 1$. It means that typically $\omega < \omega_i < T$. Since we are interested in $\epsilon \sim \Gamma(\omega) \ll \Gamma(\omega_i)$, then $\epsilon$-dependence in $J_1$ and thus in $\Gamma(\omega,\epsilon)$ is indeed negligible, and our assumption is justified.

We recover most singular term for logarithmic integral by expanding $f(z) = \coth (z/2) \approx 2/z$ and arrive at following equation:
\begin{equation}
\gamma(z)\approx\frac{3}{8\pi^{6}}\int_{|z_i|\lesssim 1}\frac{dz_i}{\gamma(z_{1})z_{1}^{2}+\gamma(z_{2})z_{2}^{2}+\gamma(z_{3})z_{3}^{2}}
\end{equation}
Since for constant $\gamma(z)$ the integral is logarithmic, we substitute asymptotic behavior of the form $\gamma(z \ll 1) = C_1 \sqrt{\ln\frac{1}{|z|}}$ and arrive at following expression for constant $C_1 = \sqrt{\sqrt{3} / 2\pi^5} \approx 5.3 \cdot 10^{-2}$. This gives us final result for decay rate of particles near the mass shell $\omega = u q$:
\begin{equation}
\label{eq:DecayZeroField}
\Gamma(\omega)=C_{1}\frac{\alpha\lambda_{-}}{u^{3}}T\omega^{2}\sqrt{\ln\frac{T}{|\omega|}} \sim T \frac{\omega^2}{J^2} \sqrt{\ln\frac{T}{|\omega|}}
\end{equation}
This is one of our major results in this paper.

\subsubsection{Nonzero magnetic field}
Let us now switch to the self-consistent calculation of decay rate of quasiparticles in the presence of magnetic field. Corresponding self-consistent Dyson equation for $\Gamma(\omega, \epsilon)$ is written as follows:
\begin{equation}
\Gamma(\omega,\epsilon)=\frac{\alpha_{1}^{2}q}{16\pi^{3}u^{3}}\int d\omega_i  \omega_{1}\omega_{2} J_2(\epsilon, \omega_i) (f(\omega_{1})+f(\omega_{2}))
\end{equation}
\begin{equation}
J_2(\epsilon, \omega_i)= \int d\epsilon_i \frac{1}{\pi^{2}}\frac{\Gamma(\omega_{1},\epsilon_1)\Gamma(\omega_{2},\epsilon_2)}{\left[\epsilon_{1}^{2}+\Gamma^{2}(\omega_{1},\epsilon_{1})\right]\left[\epsilon_{2}^{2}+\Gamma^{2}(\omega_{2},\epsilon_{2})\right]}
\end{equation}
Here integration is performed over all $\omega_i$ and $\epsilon_i$ with constraints $\omega_1 + \omega_2 = \omega$ and $\epsilon_1 + \epsilon_2 = \epsilon$.
Making the assumption of negligible $\epsilon$-dependence of $\Gamma(\omega,\epsilon)$, we perform $\epsilon_i$ integration
\begin{equation}
J_2(\epsilon,\omega_i) = \frac{1}{\pi} \frac{\Gamma(\omega_1) + \Gamma(\omega_2)}{\epsilon^2 + (\Gamma(\omega_1) + \Gamma(\omega_2))^2},
\end{equation}
and switch to dimensionless decay rate, which is defined as follows:  $\Gamma(\omega)=\frac{|\alpha_{1}|}{u^{2}}\omega^{2}\gamma(z \equiv \frac{\omega}{T})$. The self-consistent dimensionless integral equation for $\gamma(z)$ reads as follows:
\begin{equation}
\gamma(z)=\frac{1}{16\pi^{4}}\frac{1}{z}\int dz_i\frac{z_{1}z_{2}}{z_{1}^{2}\gamma(z_{1})+z_{2}^{2}\gamma(z_{2})}(f(z_{1})+f(z_{2}))
\end{equation}
Similarly to the previous discussion, we first assume that $\gamma(z)$ is slowly varying function of $z$ for small $z$; after making this assumption for the r.h.s. of the equation, we immediately recover $1/z$ behavior of the integral, and main contribution now comes from the area $z_i \sim z \ll 1$. That again allows us to expand distribution functions $f(z) \approx 2/z$ and to write the following equation for $\gamma(z \ll 1)$:
\begin{equation}
\gamma(z)=\frac{1}{8\pi^{4}}\int\frac{dz_i}{z_{1}^{2}\gamma(z_{1})+z_{2}^{2}\gamma(z_{2})}
\end{equation}

Finally, substituting $\gamma(z) = C_2 / \sqrt{|z|}$ and performing integration, we evaluate the value of $C_2$:
\begin{equation}
C_2^2 = \frac{1}{8\pi^{4}}\int_{-\infty}^{\infty}\frac{dx}{|x|^{3/2}+|1-x|^{3/2}},\quad C_2 \approx 7.08 \cdot 10^{-2},
\end{equation}
which yields final result for decay rate for this case:
\begin{equation}
\label{eq:DecayFiniteField}
\Gamma(\omega)=C_2 \frac{|\alpha_{1}|}{u^{2}}T^{1/2}|\omega|^{3/2} \sim \frac{|h| T^{1/2} |\omega|^{3/2}}{J^2} 
\end{equation}

Let us now switch to the discussion of the obtained result. The main contribution to the decay rate comes from the the modes with $\omega_i \sim \omega \ll T$. Surprisingly, these modes correspond to the classical limit; indeed, the result was originally obtained for \textit{classical one-dimensional liquid} in Ref. \cite{Andreev}; the same result was more recently rederived within Luttinger Liquid
framework~\cite{Samokhin}.

However, we obtained that main contribution now comes from $\omega_i \sim \omega$, which is inconsistent with original assumption of negligible $\epsilon$-dependence. We argue below that taking $\epsilon$-dependence into the account will lead to the same behavior of the linewidth
 $\Gamma(\omega) \propto T^{1/2} |\omega|^{3/2}$, but with modified constant $C_2$.

\subsubsection{Crossover region}
We have calculated  decay  rate of  quasiparticles for zero and relatively large magnetic fields, Eqs. \eqref{eq:DecayZeroField} and \eqref{eq:DecayFiniteField} respectively. For the validity of  self-consistent procedure performed above, either one or another decay
mechanism should be parametrically stronger than another.

For a particle with an arbitrary energy $\omega \ll T$, there is a crossover magnetic field $h_c(\omega)$ where two contributions for the decay rate are almost equal; namely, comparing the results, we immediately obtain $h_c(\omega) \propto \sqrt{\omega T \ln \frac{T}{\omega}}$. The crossover between answers \eqref{eq:DecayZeroField} and \eqref{eq:DecayFiniteField} happens at that threshold magnetic field $h_c(\omega)$, below which the first answer can be applied, and above which the second one is correct.
At $h \gg h_c(T) \sim T$ all the excitations ``feel'' the magnetic field.

\subsection{Higher diagrams}
In the previous section we have developed a self-consistent approach, which corresponded to the resummation of the infinite series of Feynman diagrams. However, a priori it is not known which diagrams mostly contribute to the width of the spectral weight, thus one needs to perform a sanity check. However all the virtual particles participating in the processes contributing to the most singular part of 
of the self-energy of right-moving plasmon are  right-movers as well.

In this section we will consider diagrams coming from the next order of perturbation theory in order to see if they contain smallness compared to ones already taken into account.

\subsubsection{Zero magnetic field}
\begin{figure}[ht]
	\includegraphics[width=0.5\columnwidth]{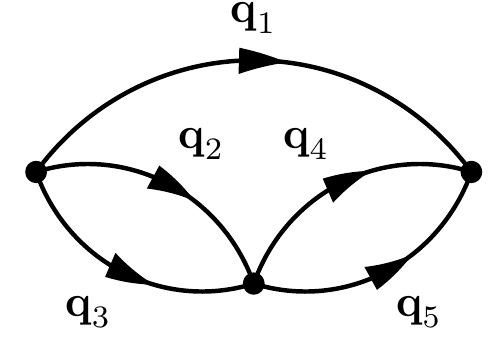}
	\caption{Next order Feynman diagram for $\Sigma^{(R)}_{ret}(\mathbf{q})$ contributing to the $\Gamma(\omega)$ for the zero magnetic field case. Bold lines correspond to dressed Keldysh Green functions for right-moving plasmons (with retarded component given by \eqref{eq:DressedGreenFunction}).}
	\label{fig:SigmaNextOrder:4}
\end{figure}

The next diagram comes from third order of perturbation theory and is shown at Fig. \ref{fig:SigmaNextOrder:4}. All the particles are right-moving, and below we will drop the $R$ superscript for all the Green functions. This diagram yields the following analytic expression:
\begin{widetext}
\begin{multline}
\Sigma_{ret}^{(3)}(\mathbf{q})=54i\alpha^{3}\lambda_{-}^{3}\int\frac{d^{2}\mathbf{q}_{i}}{(2\pi)^{2}}\Big(g_{ret}(\mathbf{q}_{1})g_{K}(\mathbf{q}_{2})g_{K}(\mathbf{q}_{3})\left[g_{K}(\mathbf{q}_{4})g_{adv}(\mathbf{q}_{5})+g_{adv}(\mathbf{q}_{4})g_{K}(\mathbf{q}_{5})\right]+\\
+g_{ret}(\mathbf{q}_{1})\left[g_{K}(\mathbf{q}_{2})g_{ret}(\mathbf{q}_{3})+g_{ret}(\mathbf{q}_{2})g_{K}(\mathbf{q}_{3})\right]g_{ret}(\mathbf{q}_{4})g_{ret}(\mathbf{q}_{5})+\\
+g_{ret}(\mathbf{q}_{1})\left[g_{K}(\mathbf{q}_{2})g_{ret}(\mathbf{q}_{3})+g_{ret}(\mathbf{q}_{2})g_{K}(\mathbf{q}_{3})\right]g_{adv}(\mathbf{q}_{4})g_{adv}(\mathbf{q}_{5})+\\
+g_{ret}(\mathbf{q}_{1})\left[g_{K}(\mathbf{q}_{2})g_{ret}(\mathbf{q}_{3})+g_{ret}(\mathbf{q}_{2})g_{K}(\mathbf{q}_{3})\right]g_{K}(\mathbf{q}_{4})g_{K}(\mathbf{q}_{5})+\\
+g_{ret}(\mathbf{q}_{1})g_{ret}(\mathbf{q}_{2})g_{ret}(\mathbf{q}_{3})\left[g_{adv}(\mathbf{q}_{4})g_{K}(\mathbf{q}_{5})+g_{K}(\mathbf{q}_{4})g_{adv}(\mathbf{q}_{5})\right]+\\
+g_{K}(\mathbf{q}_{1})\left[g_{K}(\mathbf{q}_{2})g_{ret}(\mathbf{q}_{3})+g_{ret}(\mathbf{q}_{2})g_{K}(\mathbf{q}_{3})\right]\left[g_{K}(\mathbf{q}_{4})g_{ret}(\mathbf{q}_{5})+g_{ret}(\mathbf{q}_{4})g_{K}(\mathbf{q}_{5})\right]\Big)
\end{multline}
\end{widetext}
Here we integrate over all $\mathbf{q}_i = (q_i, \omega_i)$ with constraints $\mathbf{q}_1 + \mathbf{q}_2 + \mathbf{q}_3 = \mathbf{q}_1 + \mathbf{q}_4 + \mathbf{q}_5 = \mathbf{q}$. Substituting equilibrium relation for Keldysh Green function \eqref{eq:KeldyshEquilibrium}, and discarding some terms due to causality, we arrive at following expression (here $f_i \equiv f(\omega_i)$):
\begin{multline}
\Sigma_{ret}^{(3)}(\mathbf{q})\simeq 54i\alpha^{3}\lambda_{-}^{3}\int\frac{d^{2}\mathbf{q}_{i}}{(2\pi)^{2}}\times\\
\times\Big(g_{ret}(\mathbf{q}_{1})g_{ret}(\mathbf{q}_{2})g_{ret}(\mathbf{q}_{3})g_{adv}(\mathbf{q}_{4})g_{adv}(\mathbf{q}_{5})\times\\
\times\left((f_{2}+f_{3})(1+f_{4}f_{5})-(f_{4}+f_{5})(1+f_{2}f_{3})\right)+\\
+g_{ret}(\mathbf{q}_{1})g_{ret}(\mathbf{q}_{2})g_{ret}(\mathbf{q}_{3})g_{ret}(\mathbf{q}_{4})g_{ret}(\mathbf{q}_{5})\times\\
\times(f_{2}+f_{3})\left((1+f_{4}f_{5})+f_{1}(f_{4}+f_{5})\right)\Big).
\end{multline}
Assuming again that major contribution  comes from classical modes with $\omega \ll T$, we replace $f(\omega) \approx 2 T / \omega$. The first term cancels out in the leading order in $T / \omega$. Finally, substituting dressed Green functions and replacing $q_i$ by $\omega_i / u$ in the prefactors, we arrive at (here $\Gamma_i \equiv \Gamma(\omega_i)$)
\begin{multline}
\Sigma_{ret}^{(3)}(\omega=uq)\simeq\frac{432\alpha^{3}\lambda_{-}^{3}T{}^{3}}{(2\pi)^{5}u^{8}}\omega\times\\
\times\int\frac{d\omega_{i}}{2\pi}\frac{\omega_{2}+\omega_{3}}{(\Gamma_1+\Gamma_2+\Gamma_3)(\Gamma_1+\Gamma_4+\Gamma_5)}.
\end{multline}
The above result is purely real and corresponds to the renormalization of the spectrum instead of the linewidth which we are interested in. 
It happens that the  \textit{most singular} contribution to  the diagram  with three vertices does not contain imaginary part.
In other terms,  contribution to $\Im\Sigma$ from that diagram is parametrically 
smaller than the value \eqref{eq:DecayZeroField}.

We now proceed with the calculation. Simple power counting shows that this integral is again logarithmic (we integrate over 3 different $\omega_i$ out of 5, since there are two independent conservation laws). However, the logarithmically divergent part cancels out because it is odd in $\omega$. This cancellation means that contribution to $\Gamma(\omega)$ comes from modes $\omega_i \sim \omega$, which allows us to replace logarithms in $\Gamma(\omega_i)$ by a constant, $\ln \frac{T}{\omega_i} \mapsto \ln \frac{T}{\omega}$. Finally, substituting the result \eqref{eq:DecayZeroField}, we arrive at:
\begin{equation}
\label{eq:SigmaZeroField:correction}
\Sigma_{ret}^{(3)}(\omega)=C_1^\prime\frac{\alpha\lambda_{-}}{u^{2}}\frac{T\omega}{\ln\frac{T}{|\omega|}}
\end{equation}
with numerical constant 
\begin{equation}
\label{eq:C1prime}
C_{1}^{\prime}=\frac{9\sqrt{2}}{16\pi^{5}C_{1}^{2}}=\frac{3\sqrt{6}}{8}\approx0.91
\end{equation}

The result given by Eqs.\eqref{eq:SigmaZeroField:correction}, \eqref{eq:C1prime} of the same order in the ``coupling constant'' $\alpha \lambda_{-}$ as $\Gamma(\omega)$ obtained from the lowest-order self-consistent solution, Eq.\eqref{eq:DecayZeroField}, which is what one should expect. However, it is still parametrically smaller by the factor $\ln^{-3/2} \frac{T}{|\omega|} \ll 1$. 

Higher-order diagrams (for example, coming from fourth order of perturbation theory) still contain singular contributions to the linewidth, that are of the same order in ``coupling constant'' $\alpha \lambda_{-}$ as $\Gamma(\omega)$ coming from the first diagram, Eq. \eqref{eq:DecayZeroField}. Explicit calculation of the higher order diagrams is cumbersome, but we argue that (similarly to the third order diagram discussed above)  it won't contain logarithmically large factors and thus will be parametrically smaller by some negative power 
of $\ln \frac{T}{|\omega|} \gg 1$.

\subsubsection{Nonzero magnetic field}
\begin{figure}[ht]
	\includegraphics[width=0.5\columnwidth]{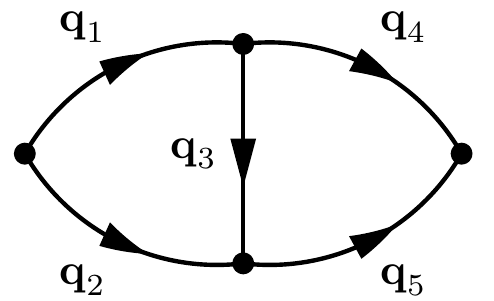}
	\caption{Next order Feynman diagram for $\Sigma^{(R)}(\mathbf{q})$ contributing to the $\Gamma(\omega)$ for nonzero magnetic field case.}
	\label{fig:SigmaNextOrder:3}
\end{figure}
First diagram which wasn't taken into account by self-consistent procedure comes from fourth order of perturbation theory and is shown on Fig. \ref{fig:SigmaNextOrder:4}. The corresponding expression for retarded self-energy yields:
\begin{widetext}
\begin{multline}
\Sigma_{ret}^{(4)}(\mathbf{q})=-2\alpha_{1}^{4}\int\frac{d^{2}\mathbf{q}_{i}}{(2\pi)^{2}} \Big(g_{ret}(\mathbf{q}_{1})g_{K}(\mathbf{q}_{2})g_{K}(\mathbf{q}_{3})g_{ret}(\mathbf{q}_{4})g_{adv}(\mathbf{q}_{5})
+g_{K}(\mathbf{q}_{1})g_{ret}(\mathbf{q}_{2})g_{adv}(\mathbf{q}_{3})g_{ret}(\mathbf{q}_{4})g_{K}(\mathbf{q}_{5})+\\
+g_{K}(\mathbf{q}_{1})g_{ret}(\mathbf{q}_{2})g_{adv}(\mathbf{q}_{3})g_{K}(\mathbf{q}_{4})g_{ret}(\mathbf{q}_{5})
+g_{ret}(\mathbf{q}_{1})g_{ret}(\mathbf{q}_{2})g_{ret}(\mathbf{q}_{3})g_{ret}(\mathbf{q}_{4})g_{adv}(\mathbf{q}_{5})+\\
+g_{ret}(\mathbf{q}_{1})g_{ret}(\mathbf{q}_{2})g_{K}(\mathbf{q}_{3})g_{ret}(\mathbf{q}_{4})g_{K}(\mathbf{q}_{5})
+g_{K}(\mathbf{q}_{1})g_{ret}(\mathbf{q}_{2})g_{K}(\mathbf{q}_{3})g_{adv}(\mathbf{q}_{4})g_{ret}(\mathbf{q}_{5})+\\
+g_{ret}(\mathbf{q}_{1})g_{K}(\mathbf{q}_{2})g_{ret}(\mathbf{q}_{3})g_{K}(\mathbf{q}_{4})g_{ret}(\mathbf{q}_{5})
+g_{ret}(\mathbf{q}_{1})g_{K}(\mathbf{q}_{2})g_{ret}(\mathbf{q}_{3})g_{ret}(\mathbf{q}_{4})g_{K}(\mathbf{q}_{5})+\\
+g_{ret}(\mathbf{q}_{1})g_{ret}(\mathbf{q}_{2})g_{adv}(\mathbf{q}_{3})g_{adv}(\mathbf{q}_{4})g_{ret}(\mathbf{q}_{5})
+g_{ret}(\mathbf{q}_{1})g_{ret}(\mathbf{q}_{2})g_{K}(\mathbf{q}_{3})g_{K}(\mathbf{q}_{4})g_{ret}(\mathbf{q}_{5})\Big).
\end{multline}
\end{widetext}
Here we integrate over all the $\mathbf{q}_i = (q_i, \omega_i)$, that are not fixed by the energy-momentum conservation laws, which are as follows: $\mathbf{q}_1 + \mathbf{q}_2 = \mathbf{q}_4 + \mathbf{q}_5 = \mathbf{q}$, $\mathbf{q}_1 = \mathbf{q}_3 + \mathbf{q}_4$ (thus only two energy-momentum pairs are independent). Using equilibrium relation \eqref{eq:KeldyshEquilibrium} and noticing that some terms vanish due to retarded structure of corresponding diagram, we regroup these terms as follows (here again $f_i \equiv f(\omega_i)$):
\begin{widetext}
\begin{multline}
\Sigma_{ret}^{(4)}(\mathbf{q})=-2\alpha_{1}^{4}\int\frac{d^{2}\mathbf{q}_{i}}{(2\pi)^{2}}\Big(g_{ret}(\mathbf{q}_{1})g_{ret}(\mathbf{q}_{2})g_{ret}(\mathbf{q}_{3})g_{ret}(\mathbf{q}_{4})g_{adv}(\mathbf{q}_{5})(1+f_{2}f_{3}-f_{3}f_{5}-f_{2}f_{5})+\\
+g_{ret}(\mathbf{q}_{1})g_{ret}(\mathbf{q}_{2})g_{adv}(\mathbf{q}_{3})g_{adv}(\mathbf{q}_{4})g_{ret}(\mathbf{q}_{5})(1+f_{3}f_{4}-f_{1}f_{4}-f_{1}f_{3})+\\
+g_{ret}(\mathbf{q}_{1})g_{ret}(\mathbf{q}_{2})g_{adv}(\mathbf{q}_{3})g_{ret}(\mathbf{q}_{4})g_{ret}(\mathbf{q}_{5})(f_{1}-f_{3})(f_{4}+f_{5})+\\
+g_{ret}(\mathbf{q}_{1})g_{ret}(\mathbf{q}_{2})g_{ret}(\mathbf{q}_{3})g_{ret}(\mathbf{q}_{4})g_{ret}(\mathbf{q}_{5})(f_{2}+f_{3})(f_{4}+f_{5})\Big).
\end{multline}
\end{widetext}

We again apply assumption that main contribution comes from classical modes and thus $f(\omega) \approx 2 T / \omega$ and immediately obtain that first two terms vanishes in the leading order. Next we substitute dressed Green functions \eqref{eq:DressedGreenFunction} and integrate over $q_i$, neglecting the $\epsilon$-dependence of $\Gamma(\omega,\epsilon)$, which was discussed above for the sake of comparison:
\begin{multline}
\Sigma_{ret}^{(4)}(\omega = u q)=i\frac{8\alpha_{1}^{4}T^{2}}{(2\pi)^{5}u^{7}}\omega\int\frac{d\omega_{i}}{2\pi} \times\\
\times\frac{1}{(\Gamma_1 + \Gamma_2)(\Gamma_4 + \Gamma_5)}\Big(\frac{\omega_{2}\omega_{4}}{\Gamma_{1}+\Gamma_{3}+\Gamma_{5}}+\frac{\omega_{1}\omega_{5}}{\Gamma_{2}+\Gamma_{3}+\Gamma_{4}}\Big)
\end{multline}
Substitution of \eqref{eq:DecayFiniteField} yields following expression for the correction:
\begin{equation}
\delta \Gamma(\omega) = -C_2^\prime\frac{|\alpha_{1}|T^{1/2}}{u^{2}}|\omega|^{3/2}.
\end{equation}
with numerical constant estimated as $C_{2}^{\prime}\approx1.6\cdot10^{-2}$. One finds that the correction is of the same order as the
lowest-order result itself, has the different sign and numerical constant that is almost 4 times smaller.  Contrary to the case 
$h=0$ studied before, here all relevant excitation energies $\omega_i \sim \omega $ and all diagrams are of the same order of magnitude.
It was noticed in Ref.~\cite{KPZrelation} that this problem
appears to be asymptotically  equivalent to the Kardar-Parisi-Zhang nonlinear model~\cite{KPZ} of  noisy classical dynamics.

\section{Correlation functions}
\label{sec:CorrelationFunctions}
We have calculated the decay rates of the quasiparticles on the mass shell. In this Section we will discuss how this decay rates manifest themselves in the real-space properties of the spin-spin correlation functions, i.e. $\left<\hat{S}^z_{x_1}(t_1) \hat{S}^z_{x_2}(t_2)\right>$. This correlation is equivalent (via the Jordan-Wigner transformation) to the fermionic density-density correlation function, which in turn is related to the plasmon propagator. To be more specific:
\begin{multline}
\left<\hat{S}^z(x_1, t_1) \hat{S}^z(x_2, t_2)\right> = \\
= K a^2 \left<(R+L)(x_1,t_1)(R + L)(x_2,t_2)\right> = \\
= K a^2 (g_{<}^{(R)}(\mathbf{x}_1-\mathbf{x}_2) + g_{<}^{(L)}(\mathbf{x}_1-\mathbf{x}_2)),
\end{multline}
with $\mathbf{x}_i = (x_i, t_i)$ and $a$ being lattice constant. 
In the above expression we discarded the offdiagonal Green functions $\left<R L\right>_{ret}$, which appear in the presence of the magnetic field; as it was discussed in Sec. \ref{sec:LowestOrder}, they are parametrically smaller and thus negligible. Correlation functions for left- and right-moving plasmons are related via plasmon velocity sign change $u \mapsto -u$. Below we will focus on the calculating $g_{<}^{(R)}(x,t)$.

Without take into account the effects of spectrum nonlinearity, one arrives at $\Im g^{(R)}_{ret} = -\frac{q}{2} \delta(\omega - u q)$. Inverting Fourier transform give us well-known pure LL result \cite{GiamarchiBook}:
\begin{equation}
\label{eq:GreenPureLL}
g^{(R)}_{<}(x,t) = \frac{1}{4\pi^{2}}\cdot\frac{\pi^2 T^2}{u^2\sinh^{2}\frac{\pi T(x-ut)}{u}}
\end{equation}
This result describes a wave packet centered at $x = u t$, with a constant width $l_T \equiv \frac{u}{T}$.

Above we have evaluated the decay rate for $\omega \ll T$. This leads to the modification of \eqref{eq:GreenPureLL}.
Direct calculation (see Appendix \ref{sec:AppendixCorrelation} for details) shows that modified answer behaves as quasi-diffusive wave packet, which is still centered at $x = u t$, but its width now depends on time as 
\begin{equation}
\label{eq:DiffusiveLengthscale}
l_D = \begin{cases}
(C_{1}\alpha\lambda_{-}T|t|/u)^{1/2}, & h\ll T\\
(T/u)^{1/3}(C_{2}\alpha_{1}|t|)^{2/3}, & h\gg T
\end{cases}
\end{equation}
which can be estimated using the original model parameters as
\begin{equation}
l_{D} \sim\begin{cases}
a\cdot(T|t|)^{1/2}, & h\ll T\\
a\cdot(T/J)^{1/3}(h|t|)^{2/3}, & h\gg T
\end{cases}
\end{equation}
It is also convenient to introduce dimensionless distance to the ``light cone'' $\delta = (x - ut) / l_D$. For the ``height'' of the corresponding wave packet, one finds the following expression:
\begin{equation}
\label{eq:GreenRealTimeCone}
g_{<}^{(R)}(|x - u t|\ll l_D)\approx\frac{1}{l_T l_{D}}\cdot\begin{cases}
\frac{\Gamma(5/3)}{2\pi^{2}}, & h\ll T\\
\frac{1}{4\pi^{3/2}\ln^{1/4}(l_{D} / l_T)}, & h\gg T
\end{cases}
\end{equation}
while its tails behave in a following manner:
\begin{multline}
\label{eq:GreenRealTimeTail}
g_{<}^{(R)}(|x| \gg |x - ut| \gg l_D) \approx \\ 
\approx \frac{1}{l_{T} l_{D}}\begin{cases}
\frac{1}{4\pi^{3/2}\ln^{1/4}(\delta l_{D}/l_{T})}\exp\left(-\frac{\delta^{2}}{4\ln^{1/4}(\delta l_{D}/l_{T})}\right), & h\ll T\\
\frac{3}{8\sqrt{2}\pi^{3/2}}\frac{1}{|\delta|^{5/2}}, & h\gg T
\end{cases}
\end{multline}
These asymptotics work only while $l_D(t) \gg l_T$.

This leads to the following answers for the spin-spin correlation functions. At the ``light cone'' (that is $\quad|n\mp ut/a|\ll l_{D}/a$), for zero magnetic field one has:
\begin{equation}
\left\langle S_{n}^{z}(t)S_{0}^{z}(0)\right\rangle =\frac{1}{4\pi^{3/2}}\frac{a^{2}}{l_{T}l_{D}}\frac{1}{\ln^{1/4}(l_{D}/l_{T})}
\end{equation}
while for nonzero magnetic field one has:
\begin{equation}
\left\langle S_{n}^{z}(t)S_{0}^{z}(0)\right\rangle \approx\frac{\Gamma(5/3)}{2\pi^{2}}\cdot\frac{a^{2}}{l_{T}l_{D}}.
\end{equation}
Away from light cone (that is $|n\mp ut/a|\gg l_{D}/a$), for zero magnetic field one has:
\begin{multline}
\left\langle S_{n}^{z}(t)S_{0}^{z}(0)\right\rangle =\frac{1}{4\pi^{3/2}}\frac{a^{2}}{l_{T}l_{D}}\ln^{-1/4}\left(\frac{|na\mp ut|}{l_{T}}\right)\times\\
\times\exp\left(-\frac{1}{4}\frac{(na\mp ut)^{2}}{l_{D}^{2}}\ln^{-1/4}\left(\frac{|na\mp ut|}{l_{T}}\right)\right),
\end{multline}
while for nonzero magnetic field one has:
\begin{equation}
\left\langle S_{n}^{z}(t)S_{0}^{z}(0)\right\rangle \approx\frac{3}{8\sqrt{2}\pi^{3/2}}\frac{l_{D}^{3/2}}{l_{T}a^{1/2}}\frac{1}{|n\mp ut/a|^{5/2}}
\end{equation}

\section{Conclusions}
\label{sec:Conclusions}
In this paper we calculated  dynamic  spin-spin correlation $\left<\hat{S}^z(x_1, t_1) \hat{S}^z(x_2, t_2)\right>$
for the XXZ spin chain with ferromagnetic $z-z$ coupling, $\Delta >0$, in the semiclassical regime where  
relevant frequency $\omega$ is much smaller than the temperature $T$. 
The results are given by Eqs.\eqref{eq:DecayZeroField}, \eqref{eq:DecayFiniteField} for the limiting cases of  small $h \ll T$ and large $h \gg T$
values of magnetic field $h$. Physically these results describe motion and spreading of a "wave packet" of a spin
excitation due to scattering by thermal excitations of the same spin chain.

The key quantity needed to derive these results is a decay rate $\Gamma(\omega)$ of bosonic quasiparticle with 
energy $\omega$, which appears in the Luttinger Liquid description of the  interacting fermionic problem
equivalent to the original XXZ model via Jordan-Wigner transformation.  Lifetime of bosonic modes is finite
due to nonlinear coupling between them.  In turn, this coupling is due to Fermionic spectrum nonlinearity 
at the energies close to the Fermi level. 

If  original problem spin is symmetric with respect to 
 $S_z \to - S_z$ reflection, the particle-hole symmetry is preserved in the Fermionic representation; 
as a result, the lowest-order nonlinearity in the spectrum of Fermions is cubic, 
$\delta\epsilon(q) \propto q^3$. After bozonization, it translates into the 4-order nonlinear coupling between 
bosonic modes.  We have calculated the resulting $\Gamma(\omega)$   at  $\omega \ll T$ by means of 
self-consistent summation of the main set of diagrams, the result is provided in Eq.(\ref{eq:DecayZeroField}).
It was shown that contributions from higher diagrams is smaller as negative powers of $\ln\frac{T}{\omega}$.

In presense of nonzero magnetic field $h$ the above-mentioned symmetry is broken and cubic nonlinear vertices
are present in the bosonic representation.  Relative importance of these cubic terms and 4-order terms
for the decay rate $\Gamma(\omega)$  depends on all three parameters  $\omega$, $T$ and $h$. Namely,
cubic nonlinearity dominates at $h^2 \gg \omega T \ln\frac{T}{\omega}$.  Our results for $\Gamma(\omega)$
in this region coinside (up to numeric coefficient of order 1) with those previously obtained 
in~\cite{Andreev,Samokhin}. 
Applicability range of these results was questioned recently in 
Ref.~\cite{Gangardt}, where the statement was made that "hydrodynamic" scaling behavior 
(\ref{eq:DecayFiniteField}) is limited to a very narrow frequency range
$\omega \leq \omega^* = 1/\tau(T) \propto T^7$, while at higher frequencies $\Gamma(\omega) \sim 1/\tau (T)$,
see Fig.1 of the paper~\cite{Gangardt}. 
We doubt this statement is correct since direct substitution of $\omega = 1/\tau(T)$ into Eq.(\ref{eq:DecayFiniteField}) 
leads to the estimate for $\Gamma (1/\tau (T)) \sim T^{11} \ll 1/\tau (T)$. Thus we tend to believe that the issue
of applicability of the result (\ref{eq:DecayFiniteField}) is still open.

We are grateful to  I. V. Gornyi,  A. D. Mirlin, I. V. Protopopov, M. A. Skvortsov for illuminating discussions.

This research was supported by  the Russian Science Foundation grant
\#  14-42-00044.
The research was also partially supported by the RF Presidential Grant No.\ NSh-10129.2016.2.

\appendix
\section{Calculation of the imaginary part of the self-energy}
\label{sec:AppendixImSigma}
In this Appendix we will demonstrate how to obtain the expressions for the imaginary parts of the self-energy. 

\paragraph{Zero magnetic field}
We start from the analytical expression for the two diagrams on the Fig. \ref{fig:selfenergy4}, that are given by the second order of perturbation theory (here $\textbf{q}_3 = \textbf{q} - \textbf{q}_1 - \textbf{q}_2$):
\begin{multline}
\label{eq:SigmaRto3R}
\Sigma_{ret}^{(R \to 3R)}(\mathbf{q})=-6\alpha^{2}\lambda_{-}^{2}\int\frac{d^{2}\mathbf{q}_{1}}{(2\pi)^{2}}\frac{d^{2}\mathbf{q}_{2}}{(2\pi)^{2}} \times \\
\times\Big[g_{ret}^{(R)}(\mathbf{q}_{1})g_{ret}^{(R)}(\mathbf{q}_{2})g_{ret}^{(R)}(\mathbf{q}_{3})+g_{ret}^{(R)}(\mathbf{q}_{1})g_{K}^{(R)}(\mathbf{q}_{2})g_{K}^{(R)}(\mathbf{q}_{3})+\\
+g_{K}^{(R)}(\mathbf{q}_{1})g_{ret}^{(R)}(\mathbf{q}_{2})g_{K}^{(R)}(\mathbf{q}_{3})+g_{K}^{(R)}(\mathbf{q}_{1})g_{K}^{(R)}(\mathbf{q}_{2})g_{ret}^{(R)}(\mathbf{q}_{3})\Big]
\end{multline}
\begin{multline}
\label{eq:SigmaRtoRplus2L}
\Sigma_{ret}^{(R\to R+2L)}(\mathbf{q})=-\frac{\alpha^{2}\lambda_{+}^{2}}{2}\int\frac{d^{2}\mathbf{q}_{1}}{(2\pi)^{2}}\frac{d^{2}\mathbf{q}_{2}}{(2\pi)^{2}} \times \\
\times\Big[g_{ret}^{(R)}(\mathbf{q}_{1})g_{ret}^{(L)}(\mathbf{q}_{2})g_{ret}^{(L)}(\mathbf{q}_{3})+g_{ret}^{(R)}(\mathbf{q}_{1})g_{K}^{(L)}(\mathbf{q}_{2})g_{K}^{(L)}(\mathbf{q}_{3})+\\
+g_{K}^{(R)}(\mathbf{q}_{1})g_{ret}^{(L)}(\mathbf{q}_{2})g_{K}^{(L)}(\mathbf{q}_{3})+g_{K}^{(R)}(\mathbf{q}_{1})g_{K}^{(L)}(\mathbf{q}_{2})g_{ret}^{(L)}(\mathbf{q}_{3})\Big]
\end{multline}

We proceed by expressing retarded Greens functions through their imaginary parts using Kramers-Kronig relations $g_{ret}(\omega)=\frac{1}{\pi}\int d\omega^{\prime}\frac{{\rm Im}g_{ret}(\omega^{\prime})}{\omega^{\prime}-\omega-i0}$; we also express Keldysh Greens functions using equilibrium relation \eqref{eq:KeldyshEquilibrium}. Next we perform integration over $\omega_i$ using known positions of residues (namely, $\omega_i = \omega_i^{\prime} - i0$). Finally, taking the imaginary part of the obtained expression using $\Im \frac{1}{x - i0} = \pi \delta(x)$, we arrive at following expressions:
\begin{multline}
{\rm Im}\Sigma_{ret}^{(R\to 3R)}(\mathbf{q})=\frac{3\alpha^{2}\lambda_{-}^{2}}{2\pi^{4}}\int d^{2}\mathbf{q}_{1}d^{2}\mathbf{q}_{2}\times\\
\times{\rm Im}g_{ret}^{(R)}(\mathbf{q}_{1}){\rm Im}g_{ret}^{(R)}(\mathbf{q}_{2}){\rm Im}g_{ret}^{(R)}(\mathbf{q}_{3})\times\\
\times(1+f(\omega_{2})f(\omega_{3})+f(\omega_{1})f(\omega_{3})+f(\omega_{1})f(\omega_{2}))
\end{multline} 
\begin{multline}
\label{eq:ImSigmaRtoRplus2L}
{\rm Im}\Sigma_{ret}^{(R\to R+2L)}(\mathbf{q})=\frac{\alpha^{2}\lambda_{+}^{2}}{8\pi^{4}}\int d^{2}\mathbf{q}_{1}d^{2}\mathbf{q}_{2} \times \\
\times {\rm Im}g_{ret}^{(R)}(\mathbf{q}_{1}){\rm Im}g_{ret}^{(L)}(\mathbf{q}_{2}){\rm Im}g_{ret}^{(L)}(\mathbf{q}_{3}) \times\\
\times(1+f(\omega_{2})f(\omega_{3})+f(\omega_{1})f(\omega_{3})+f(\omega_{1})f(\omega_{2}))
\end{multline}

For the unperturbed Greens functions \eqref{eq:ZeroGreenFunction}, the imaginary part is proportional to the delta-function at mass shell $\delta(\omega \mp uq)$, which for the \eqref{eq:ImSigmaRto3R} leads to the notorious singular behaviour of the self-energy on the mass shell:
\begin{equation}
\Im\Sigma_{ret}^{(R\to3R)}(\mathbf{q})=-\frac{3\alpha^{2}\lambda_{-}^{2}}{16\pi^{4}u^{5}}\delta(\omega-uq)I_{1}(uq,T),
\end{equation}
with 
\begin{multline}
I_{1}(\Omega,T)=\int d\omega_{1}d\omega_{2}d\omega_{3}\omega_{1}\omega_{2}\omega_{3} \delta(\omega_{1}+\omega_{2}+\omega_{3}-\Omega)\times\\
\times (1+f(\omega_{1})f(\omega_{2})+f(\omega_{1})f(\omega_{3})+f(\omega_{2})f(\omega_{3}))
\end{multline}
At zero temperature $f(\omega) = \mathrm{sign} \omega$ and the integration happens only in the region where all the $\omega_i$ of the same sign. The evaluation is then straightforward and yields $I_{1}(\Omega,T=0) =\frac{1}{30}\Omega^{5}$. The derivative w.r.t. temperature can be, however, expressed via the another integral:
\begin{equation}
\frac{\partial I_{1}}{\partial T} = 3\times\int d\omega_{1}\frac{\partial f(\omega_1)}{\partial T}\omega_{1}I_{2}(\Omega-\omega_{1},T)
\end{equation}
and
\begin{multline}
I_{2}(\Omega,T)=\int d\omega_{1}\omega_{1}(\Omega-\omega_{1})\left[f(\omega_{1})+f(\Omega-\omega_{1})\right] = \\
=\frac{1}{3}\Omega(\Omega^{2}+(2\pi T)^{2})
\end{multline}
The second integral can be evaluated using the same trick. Namely, at zero temperature it yields $I_{2}(\Omega,T=0)=\frac{1}{3}\Omega^{3}$; and the derivative w.r.t. temperature can also be evaluated exactly. Combining all together, we arrive at the expression for the first integral:
\begin{equation}
I_{1}(\Omega,T) = \frac{1}{30}\Omega\left[\Omega^{2}+(2\pi T)^{2}\right]\left[\Omega^{2}+4(2\pi T)^{2}\right]
\end{equation}
which finally yields the result \eqref{eq:ImSigmaRto3Rresult}.

The expression for the second diagram, Eq. \eqref{eq:ImSigmaRtoRplus2L} is not singular at the mass shell and thus is less interesting for us. It can be expressed via another integral:
\begin{equation}
{\rm Im}\Sigma_{ret}^{(R\to R+2L)}(\mathbf{q})=-\frac{\alpha^{2}\lambda_{+}^{2}}{64\pi^{4}}I_{3}\left(\frac{\omega - uq}{2},\frac{\omega + uq}{2},T\right)
\end{equation}
and 
\begin{multline}
I_{3}(\Omega_{1},\Omega_{2},T)=\frac{1}{2}\Omega_{1}\int d\omega_{2}\omega_{2}(\Omega_{2}-\omega_{2}) \times \\
\times(1+f(\Omega_{1})f(\omega_{2})+f(\Omega_{1})f(\Omega_{2}-\omega_{2})+f(\omega_{2})f(\Omega-\omega_{2})) = \\
= \frac{1}{2}\Omega_{1}f(\Omega_{1})I_{1}(\Omega_{2},T) + 4 \Omega_{1} T^{3} I_{4}\left(\frac{\Omega_2}{2 T}\right)
\end{multline}
\begin{equation}
I_{4}(A)=\int dz\cdot z(A-z)(1+\coth z\coth(A-z))
\end{equation}
The latter integral can be evaluated exactly only using polylogarithm functions; however, its asymptotic behavior is easily obtainable:
\begin{equation}
I_{4}(A)\approx\begin{cases}
\frac{\pi^{2}}{3}, & A\ll1\\
\frac{A^{3}}{3}, & A\gg1
\end{cases}
\end{equation}

Combining all together, we arrive at results \eqref{eq:ImSigmaRtoRplus2Lresult1} and \eqref{eq:ImSigmaRtoRplus2Lresult2}.

\paragraph{Nonzero magnetic field}
The results for cubic interaction between plasmons was studied in Ref.\cite{Aristov}. Here we will demonstrate how to reproduce it in the same fashion. We start from the expressions for two diagrams \ref{fig:selfenergy3} (here $\mathbf{q}_2 = \mathbf{q}-\mathbf{q}_1$):
\begin{multline}
\label{eq:SigmaRto2R}
\Sigma_{ret}^{(R\to2R)}(\mathbf{q})=i\alpha_{1}^{2}\int\frac{d^{2}\mathbf{q}_{1}}{(2\pi)^{2}} \times\\
\times\left[g_{K}^{(R)}(\mathbf{q}_{1})g_{ret}^{(R)}(\mathbf{q}_{2})+g_{ret}^{(R)}(\mathbf{q}_{1})g_{K}^{(R)}(\mathbf{q}_{2})\right]
\end{multline}
\begin{multline}
\label{eq:SigmaRtoRplusL}
\Sigma_{ret}^{(R\to R+L)}(\mathbf{q})=\frac{i\alpha_{2}^{2}}{2}\int\frac{d^{2}\mathbf{q}_{1}}{(2\pi)^{2}}\times\\
\times(g_{K}^{(L)}(\mathbf{q}_{1})g_{ret}^{(R)}(\mathbf{q}_{2})+g_{ret}^{(L)}(\mathbf{q}_{1})g_{K}^{(R)}(\mathbf{q}_{2})),
\end{multline}
Expressions for its imaginary parts obtained using Krammers-Kronig relations read as follows:
\begin{multline}
{\rm Im}\Sigma_{ret}^{(R\to2R)}(\mathbf{q})=-\frac{\alpha_{1}^{2}}{2\pi^{2}}\int d^{2}\mathbf{q}_{1}\times\\
\times{\rm Im}g_{ret}^{(R)}(\mathbf{q}_{1}){\rm Im}g_{ret}^{(R)}(\mathbf{q}_{2})(f(\omega_{1})+f(\omega_{2}))
\end{multline}
\begin{multline}
\label{eq:ImSigmaRtoRplusL}
{\rm Im}\Sigma_{ret}^{(R\to R+L)}(\mathbf{q})=-\frac{\alpha_{2}^{2}}{4\pi^{2}}\int d^{2}\mathbf{q}_{1}\times\\
\times{\rm Im}g_{ret}^{(L)}(\mathbf{q}_{1}){\rm Im}g_{ret}^{(R)}(\mathbf{q}_{2})(f(\omega_{1})+f(\omega_{2}))
\end{multline}
Substituting unperturbed Greens functions in the first expression, we immediately obtain that it is again singular at the mass shell, and it can expressed via the same $I_2(\Omega,T)$ integral we have evaluated above as follows:
\begin{equation}
{\rm Im}\Sigma_{ret}^{(R\to2R)}(\mathbf{q})=-\frac{\alpha_{1}^{2}}{8\pi^{2}u^{3}}\delta(\omega-uq)I_{1}(uq,T),
\end{equation}
while the second expression contains double integration over two delta-functions and is thus trivial:
\begin{multline}
{\rm Im}\Sigma_{ret}^{(R\to R+L)}(\mathbf{q})=-\frac{\alpha_{2}^{2}}{128\pi^{2}u^{3}}(\omega^{2}-u^{2}q^{2})\times\\
\times\left(f\left(\frac{\omega-uq}{2}\right)+f\left(\frac{\omega+uq}{2}\right)\right),
\end{multline}
which are precisely the results \eqref{eq:ImSigmaRto2Rresult} and \eqref{eq:ImSigmaRtoRplusLresult}.

\section{Calculation of real-space correlation functions}
\label{sec:AppendixCorrelation}
In this Appendix we will derive the results outlined in Sec. \ref{sec:CorrelationFunctions}. We need to calculate the Fourier transform of the lesser component of Keldysh Green function, which is expressed using the equilibrium relation:
\begin{equation}
\label{eq:GreaterEquilibrium}
g^{(R)}_{<}(\mathbf{q}) = \frac{2\Im g^{(R)}_{ret}(\mathbf{q})}{1 - e^{-\beta \omega}}.
\end{equation}
Using the general form of dressed Green functions \eqref{eq:DressedGreenFunction}, and neglecting the $\epsilon = \omega-u q$ dependence of decay rates $\Gamma(\omega, \epsilon)$, we can perform momentum integration and arrive at following general expression:
\begin{equation}
g_{<}^{(R)}(x,t)=\frac{T}{2\pi u^{2}}\int\frac{d\omega}{2\pi}\frac{\omega+i\Gamma{\rm sign}x}{\omega}e^{(i\omega(x-ut)-\Gamma|x|)/u}
\end{equation}

\paragraph{Nonzero magnetic field}
After substituting the result \eqref{eq:DecayFiniteField}, it is convenient to make the integral dimensionless by introducing the lengthscale $l_D$, see Eq. \eqref{eq:DiffusiveLengthscale}, $\delta = (x - u t) / l_D$ and switching to dimensionless variable $z = \omega l_D / u$. We arrive at following integral:
\begin{equation}
g_{<}^{(R)}(x,t)=\frac{1}{2\pi^{2}}\frac{1}{l_{D}l_{T}}\int_{0}^{\infty}dze^{-z^{3/2}}\left[\cos(z\delta)-\frac{l_{D}}{x}z{}^{1/2}\sin(z\delta)\right]
\end{equation}
For $\delta \ll 1$ (that is the light cone $x = u t$) the expression in the brackets can be replaced by unity. The integral is then immediately yields $\Gamma(5/3)$ (Euler gamma-function), which gives the first part of the result \eqref{eq:GreenRealTimeCone}.

For $\delta \gg 1$ one can rotate the integration contour by $\pi/2$ arriving at exponentially decaying integrals $\propto e^{-\delta z}$, and then Taylor-expand the corresponding expression, arriving at:
\begin{multline}
J_{1}(\delta)=\int_{0}^{\infty}dze^{-z^{3/2}}\cos(z\delta)= \\
= {\rm Re}\left[i\int_{0}^{\infty}dz\cdot e^{-e^{\frac{3i\pi}{4}}z^{3/2}-z\delta}\right]\approx\\
\approx\frac{1}{\sqrt{2}}\int_{0}^{\infty}dz\cdot z^{3/2}e^{-z\delta}=\frac{\Gamma(5/2)}{\sqrt{2}}\frac{1}{\delta^{5/2}},
\end{multline}
\begin{multline}
J_{2}(\delta)=\int_{0}^{\infty}dze^{-z^{3/2}}z^{1/2}\sin(z\delta)=\\
={\rm Im}\left[i\int_{0}^{\infty}dze^{-e^{3i\pi/4}z^{3/2}-z\delta}e^{i\pi/4}z^{1/2}\right]\approx\\
\approx\frac{1}{\sqrt{2}}\int_{0}^{\infty}z^{1/2}e^{-z\delta}dz=\frac{\Gamma(3/2)}{\sqrt{2}}\frac{1}{\delta^{3/2}}.
\end{multline}
Substitution of these two integrals yields the first part of the result \eqref{eq:GreenRealTimeTail}.

\paragraph{Zero magnetic field}
We again introduce lengthscale $l_D$ as in Eq. \eqref{eq:DiffusiveLengthscale}, and arrive at dimensionless integral:
\begin{multline}
g_{<}^{(R)}(x,t)=\frac{1}{2\pi^{2}}\frac{1}{l_{D}l_{T}}\int_{0}^{\infty}dze^{-z^{2}\sqrt{\ln\frac{l_{D}}{zl_{T}}}}\times\\
\times\left[\cos(z\delta)-\frac{l_{D}}{x}z\sqrt{\ln\frac{l_{D}}{zl_{T}}}\sin(z\delta)\right]
\end{multline}
For $\delta \ll 1$ one can again immediately replace the expression in the square brackets by unity, then replace the slowly-varying logarithm by constant $\ln l_D / l_T$, and then perform the Gaussian integration, which yields the value $\sqrt{\pi} / 2\ln^{1/4} l_D / l_T$. This leads us to the second part of the answer \eqref{eq:GreenRealTimeCone}.

For $\delta \gg 1$, the integrals are again almost Gaussian (after the slowly varying logarithm being replaced by its typical value):
\begin{multline}
J_{3}(\delta)=\int_{0}^{\infty}dze^{-z^{2}\sqrt{\ln(l_{D}/zl_{T})}}\cos(z\delta)\approx\\
\approx\int_{0}^{\infty}dze^{-z^{2}\sqrt{\ln(\delta l_{D}/l_{T})}}\cos(z\delta)=\\
=\frac{\sqrt{\pi}}{2\ln^{1/4}(\delta l_{D}/l_{T})}\exp\left(-\frac{\delta^{2}}{4\ln^{1/4}(\delta l_{D}/l_{T})}\right)
\end{multline}
\begin{multline}
J_{4}(\delta)=\int_{0}^{\infty}dze^{-z^{2}\sqrt{\ln(l_{D}/zl_{T})}}\sqrt{\ln(l_{D}/zl_{T})}z\sin(z\delta)\approx\\
\approx\int_{0}^{\infty}dze^{-z^{2}\sqrt{\ln(\delta l_{D}/l_{T})}}\sqrt{\ln(\delta l_{D}/l_{T})}z\sin(z\delta)=\\
=\frac{\sqrt{\pi}\delta}{4\ln^{1/4}(\beta\delta)}e^{-\frac{\delta^{2}}{4\ln^{1/4}\beta\delta}}
\end{multline}
Combining these two expression yields the second part of the answer \eqref{eq:GreenRealTimeTail}.

Let us now discuss the applicability of the obtained results for both cases of zero and nonzero magnetic fields. The results for $\Gamma(\omega)$ used here are obtained under the condition $\omega \ll T$, and thus one can use it only if the typical values of $\omega$ in all the integrals are also small compared with temperature. This criterion is equivalent to $z \ll l_D / l_T$. Since typical values of $z$ in the integrals in both cases are $z \sim 1$ (for $\delta \ll 1$) and $z \sim 1/\delta \ll 1$ (for $\delta \gg 1$), we immediately obtain that our results are applicable when $l_D \gg l_T$. This criterion corresponds to sufficiently large times.

\end{document}